\documentclass[twocolumn,trackchanges]{aastex7}

\usepackage{natbib}
\usepackage{txfonts}
\begin{document}

%\title{Absolute Calibration of Cluster Mira Variables to Anchor a $3.4\%$ Determination of the Hubble Constant}
\title{\large Large Magellanic Cloud Globular Clusters in the Near-infrared. I. RR Lyrae in Reticulum}

%% A significant change from AASTeX v6+ is in the author blocks. Now an email
%% address is required for each author. This means that each author requires
%% at least one of the following:
%%
%% \author
%% \affiliation
%% \email
%%

\author[orcid=0000-0001-6147-3360,sname='Bhardwaj']{Anupam Bhardwaj}
%%\altaffiliation{}
\affiliation{Inter-University Centre for Astronomy and Astrophysics (IUCAA), Post Bag 4, Ganeshkhind, Pune 411 007, India}
\email[show]{anupam.bhardwaj@iucaa.in}  

\author[0000-0003-3679-2428]{Susmita Das}
%%\altaffiliation{}
\affiliation{Inter-University Centre for Astronomy and Astrophysics (IUCAA), Post Bag 4, Ganeshkhind, Pune 411 007, India}
\email[]{susmita.das@iucaa.in}  

\author[]{Prashant Nishad}
%%\altaffiliation{}
\affiliation{Inter-University Centre for Astronomy and Astrophysics (IUCAA), Post Bag 4, Ganeshkhind, Pune 411 007, India}
\email[]{prashant.nishad@iucaa.in}  

\author[0000-0002-6577-2787]{Marina Rejkuba}
%%\altaffiliation{}
\affiliation{European Southern Observatory, Karl-Schwarzschild-Straße 2, 85748, Garching, Germany}
\email[]{mrejkuba@eso.org}  

\author[0000-0002-5819-3461]{Giulia De Somma}
\affiliation{INAF-Osservatorio Astronomico di Capodimonte, Via Moiariello 16, I-80131 Napoli, Italy}
\affiliation{Istituto Nazionale di Fisica Nucleare (INFN)—Sezione di Napoli, Compl. Univ. di Monte S. Angelo, Edificio G, Via Cinthia, I-80126 Napoli, Italy}
\email[]{giulia.desomma@inaf.it}  

\author[0000-0002-1330-2927]{Marcella Marconi}
\affiliation{INAF-Osservatorio Astronomico di Capodimonte, Via Moiariello 16, I-80131 Napoli, Italy}
\email[]{marcella.marconi@inaf.it}  

\author[0000-0001-8771-7554]{Chow-Choong Ngeow}
%%\altaffiliation{}
\affiliation{Graduate Institute of Astronomy, National Central University, 300 Jhongda Road, 32001 Jhongli, Taiwan}
\email[]{cngeow@gm.astro.ncu.edu.tw }  

\author[0000-0003-1801-426X]{Vincenzo Ripepi}
\affiliation{INAF-Osservatorio Astronomico di Capodimonte, Via Moiariello 16, I-80131 Napoli, Italy}
\email[]{vincenzo.ripepi@inaf.it}  

\author[]{Sarang Shah}
%%\altaffiliation{}
\affiliation{Inter-University Centre for Astronomy and Astrophysics (IUCAA), Post Bag 4, Ganeshkhind, Pune 411 007, India}
\email[]{sarang.shah@iucaa.in}  

\author[]{Subhajit Kar}
%%\altaffiliation{}
\affiliation{Inter-University Centre for Astronomy and Astrophysics (IUCAA), Post Bag 4, Ganeshkhind, Pune 411 007, India}
\email[]{subhajit.kar@iucaa.in}  

\author[]{Shashi Kanbur}
%%\altaffiliation{}
\affiliation{Department of Physics, State University of New York, Oswego, NY 13126, USA}
\email[]{shashi.kanbur@oswego.edu}  

\author[]{Matteo Monelli}
\affiliation{INAF - Osservatorio Astronomico d'Abruzzo, Teramo, Italy}
\email[]{matteo.monelli@inaf.it}  

\author[]{Massimo Dall'Ora}
\affiliation{INAF-Osservatorio Astronomico di Capodimonte, Via Moiariello 16, I-80131 Napoli, Italy}
\email[]{massimo.dallora@inaf.it}  

%\collaboration{all}{The Terra Mater collaboration}

\begin{abstract}

Reticulum is an old, metal-poor, and sparsely populated globular cluster in the outer regions of the Large Magellanic Cloud (LMC) and hosts a rich population of RR Lyrae stars. Being as close as possible to a single stellar population with negligible metallicity spread and low reddening, Reticulum is an ideal laboratory for testing stellar pulsation models and calibrating population II distance indicators. 
%Homogeneous stellar population, negligible metallicity spread, and low reddening make Reticulum an ideal laboratory for testing stellar pulsation models and calibrating population II distance indicators. 
We present homogeneous multi-epoch near-infrared (NIR, $JHK_s$) observations of RR Lyrae variables in Reticulum obtained with the Flamingos-2 imager on the 8.1-m Gemini South Telescope. Using NIR light-curve templates, we derive accurate intensity-averaged magnitudes and peak-to-peak amplitudes for 32 RR Lyrae stars, including 22 fundamental-mode (RRab), 4 first-overtone (RRc), and 6 mixed-mode (RRd) pulsators. The empirical $JHK_s$ period–luminosity (PL) relations of Reticulum RR Lyrae are very tight, exhibiting dispersions ($\sim0.05$~mag) comparable to those observed in Galactic globular clusters. The derived PL slopes are shallower than those reported for Galactic cluster variables. Adopting recent empirical and theoretical period--luminosity--metallicity (PLZ) calibrations based on Galactic globular clusters and pulsation models, we derive a true distance modulus of $\mu_0 = 18.472 \pm 0.035$ mag to Reticulum. This cluster distance is in excellent agreement with the precise geometric distance to the LMC and places Reticulum close to the LMC barycentric distance. The well-characterized RR Lyrae population and a precise distance make Reticulum a potential anchor for calibrating Population II distance ladder.
\end{abstract}

%% The AAS Journals now uses Unified Astronomy Thesaurus (UAT) 

\section{Introduction} \label{sec:intro}

RR Lyrae stars (RRLs) are fundamental tracers of old stellar populations and are among the most widely used Population II distance indicators \citep[see reviews,][]{beaton2018, bhardwaj2020}. These low-mass pulsating stars are located on the horizontal branch and are typically associated with stellar populations older than $\sim10$ Gyr \citep{catelan2009}. Consequently, RRL stars are abundant in globular clusters (GCs) and are useful probes of the formation history, chemical enrichment, and dynamical evolution of old stellar systems \citep{monelli2022}.

Resolved stellar populations in the Large Magellanic Cloud (LMC) and its GCs are particularly important for investigating the formation and evolution of the Magellanic Cloud system \cite[][]{vander2002, niederhofer2025}. Early studies of LMC GCs showed that some of the old clusters host a rich population RRL stars \citep[e.g.][]{nemec1985, walker1991} which can be used to determine distances, ages, and important insights into the metal enrichment history of the LMC. The GCs in the
Milky Way are generally divided into Oosterhoff type I (Oo-I) and II (Oo-II) with different mean periods of their RRab stars, whereas several dwarf galaxies contain Oosterhoff-intermediate (Oo-int) populations occupying the so-called ``Oosterhoff gap'' \citep{catelan2009, prudil2019, bhardwaj2022}. Several GCs that seem to fall in the Oosterhoff gap as well as the Oosterhoff types of their RR Lyrae have been linked with past merger events \citep{luongo2024}. Therefore, RRL population in GCs provides a valuable tool for tracing the formation and evolutionary history of the Milky Way halo \citep{catelan2009, prudil2024}. In this context it is notable that the LMC also contains several GCs of Oo-int RRL populations that differ significantly from those observed in Galactic GCs \citep{bono1994a, catelan2009}.

Reticulum is an old, metal-poor GC located in the outer regions of the LMC, nearly $11^\circ$ from its center \citep[\textrm{[Fe/H]}$_{ZW}\sim-1.71\pm0.1$~dex, ][]{demers1976, suntzeff1992}. Comprehensive optical time-series photometric studies of RRL stars were carried out by \citet{kuehn2011, kuehn2012, kuehn2013} in the NGC 1466, NGC 1786, and Reticulum clusters in the LMC. Using multi-epoch $BVI$ observations, \citet{kuehn2013} investigated 32 RRL stars in Reticulum and derived their photometric parameters and physical properties from Fourier decomposition of the light curves. Based on the average periods for the RRab (0.552 days) and RRc stars (0.325 days) and the ratio of number of RRc to total number of RRL (0.31), Reticulum was classified as an Oo-I cluster. 
%Similar studies by the authors identified 53 RRL stars in NGC 1786 and 49 RRL stars in NGC 1466. 
Modern large-scale variability surveys such as Optical Gravitational Lensing Experiment (OGLE) have further expanded the census of RRL stars in the LMC GCs \citep{soszynski2016}. Most of the old GCs in the LMC are metal-poor with metallicities ranging between $-1.4$ and $-2.2$~dex. \citet{sarajedini2024} derived metallicities of 14 old LMC clusters using period-amplitude-metallicity relations and found significant variations with previously published metallicity estimates in different metallicity scales.

%Despite a number of optical photometric studies of LMC GCs, near-infrared (NIR) follow-up observations have been limited to a few studies. 
Near-infrared (NIR) observations provide several important advantages over optical bands because the RRL period–luminosity (PL) relation becomes tighter at longer wavelengths and exhibits 
reduced sensitivity to reddening, metallicity, and evolutionary effects \citep[see review by][]{bhardwaj2020}. \citet{dallora2004} presented $JK_s$ observations of 30 RRL in Reticulum and derived their $K_s$-band PL relation, determining a distance to the cluster. Their distance measurement was anchored to semi-empirical relations using theoretical models and a limited sample of Hubble Space Telescope (HST) parallaxes, and the result of $18.52\pm0.11$~mag placed Reticulum at a distance beyond the LMC barycenter. In contrast, \citet{muraveva2018b} found that the cluster is located $\sim3$ kpc closer to us than the barycenter of the LMC 
based on mid-infrared (MIR) observations of RRL stars. Several other studies utilized aforementioned infrared datasets for studying RRL properties at longer wavelengths and deriving distances based on independent empirical calibrations \citep{sollima2008, braga2019, mullen2023}. Since LMC has played a critical role as primary anchor galaxy for the cosmic distance scale \citep{riess2022, casertano2026}, such NIR observations of LMC GCs offer an important independent route for improving the Population II distance scale and for investigating the structure and old stellar components of the LMC.

In this work, we present homogeneous NIR time-series observations with Gemini South Telescope for all 32 known RRL stars in Reticulum. This study is the first in a series aimed at probing variability and distance indicators in the old LMC GCs across all the NIR wavelengths. The Reticulum cluster was selected as a pilot target  because of its relatively well-studied RRL population at multiple wavelengths. Combining deep NIR photometry with existing optical information will provide improved constraints on the GC distances, reddening, and pulsation properties of their RRL population. Homogeneous RRL distances to old GCs will not only establish clusters as independent anchors to field LMC stars but also allow calibration of other population II distance indicators such as tip of the red giant branch using RRL distances \citep{beaton2016, freedman2024}. The manuscript presents description of observations and photometry in Section~\ref{sec:data}, and the details of RRL in Reticulum in Section~\ref{sec:var_rrl}. The distance determination is described in Section~\ref{sec:dis_retic} and the results are summarized in Section~\ref{sec:discuss}.

\section{Data and Photometry} \label{sec:data}

\begin{deluxetable*}{cccccccccccccc}[!t]
\tablecaption{Log of NIR observations. \label{tbl:data}}
\tabletypesize{\fontsize{8.2pt}{9.5pt}\selectfont}
\tablewidth{0pt}
\tablehead{\colhead{} &  \multicolumn{4}{c}{$J$-band}  &  \multicolumn{4}{c}{$H$-band}  &  \multicolumn{4}{c}{$K_s$-band}   & \\
\colhead{Date} & \colhead{MJD}	& \colhead{FWHM} & \colhead{Airmass} & \colhead{$N_\textrm{f}$} & \colhead{MJD}& \colhead{FWHM} & \colhead{Airmass} & \colhead{$N_\textrm{f}$} & \colhead{MJD}& \colhead{FWHM} & \colhead{Airmass} & \colhead{$N_\textrm{f}$} & ET\\
		&days	& pixel	& 	&	&days	& pixel	&  &	&days	& pixel	&   	& 	&sec}
\startdata
2023-11-30&  60278.301&     4.15&     1.38& 11&  60278.301&     3.41&     1.41& 22&  60278.301&     3.43&     1.45& 22& 10\\
2023-12-03&  60281.199&     4.27&     1.19& 11&  60281.199&     3.29&     1.25& 20&         --&       --&       --& --& 10\\
2023-12-08&  60286.102&     4.50&     1.38& 11&  60286.301&     3.10&     1.39& 22&  60286.301&     3.22&     1.43& 22& 10\\
2023-12-08&  60286.301&     3.15&     1.36& 11&         --&       --&       --& --&         --&       --&       --& --& 10\\
2023-12-11&  60289.102&     3.47&     1.39& 11&  60289.102&     3.08&     1.36& 22&  60289.102&     2.85&     1.33& 22& 10\\
2023-12-12&  60290.199&     3.69&     1.23& 11&  60290.199&     3.17&     1.24& 22&  60290.199&     3.09&     1.27& 22& 10\\
2023-12-18&  60296.102&     2.76&     1.16& 11&  60296.301&     2.57&     1.17& 22&  60296.301&     2.44&     1.18& 21& 10\\
2023-12-18&  60296.398&     3.07&     1.57& 10&  60296.398&     3.19&     1.62& 22&  60296.398&     2.75&     1.68& 23& 10\\
2023-12-24&  60302.301&     2.81&     1.27& 11&  60302.301&     2.87&     1.29& 22&  60302.301&     2.40&     1.31& 22& 10\\
2023-12-25&  60303.102&     3.37&     1.17& 11&  60303.102&     3.03&     1.16& 22&  60303.102&     2.97&     1.15& 22& 10\\
2023-12-26&  60304.102&     2.61&     1.16& 10&         --&       --&       --& --&         --&       --&       --& --& 10\\
2024-01-04&  60313.301&     3.00&     1.32& 11&  60313.301&     2.54&     1.34& 22&  60313.301&     2.52&     1.37& 22& 10\\
2024-01-07&  60316.102&     2.83&     1.15& 11&  60316.102&     2.97&     1.15& 22&  60316.102&     2.72&     1.14& 33& 10\\
\enddata
\tablecomments{MJD: Modified Julian Date (JD$-$2,400,000.5). FWHM: Measured median full width at half maximum using {\texttt SExtractor}. $N_\textrm{f}$: Number of dithered frames per epoch. ET: Exposure time (in seconds) for each dithered frame.}
\end{deluxetable*}

\begin{figure*}[!t]
\centering
\includegraphics[width=0.99\textwidth]{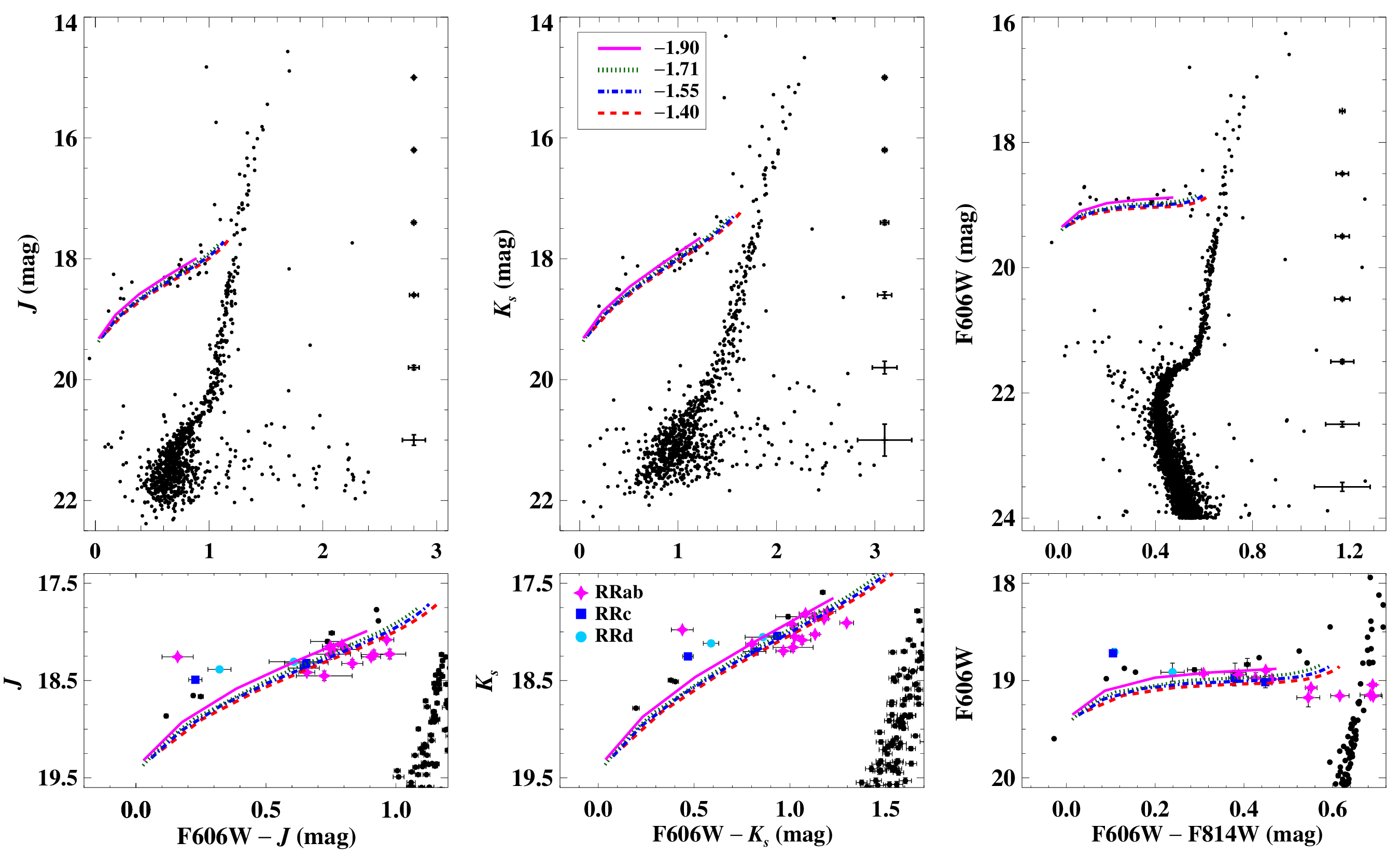}
\caption{Extinction corrected optical-NIR color-magnitude diagrams of Reticulum cluster (top panels). Representative $\pm3\sigma$ errors are also shown. The zero-age horizontal branch models with different metallicities of [Fe/H $=-1.40$ (red), $-1.55$ (blue), $-1.71$ (green), and $-1.90$ (magenta) from \citet{pietrinferni2021} are also overplotted with an adopted distance of 18.477 mag to the LMC. The bottom panels display zoomed-in view of the horizontal branch stars. Variable stars from \citet{kuehn2013} recovered within the common field of view of {\it HST} and Gemini-F2 are also shown.}
\label{fig:cmd}
\end{figure*}

\subsection{Observations and photometry}

NIR observations of Reticulum cluster were obtained using the FLAMINGOS-2 imaging spectrograph mounted on the 8.1 m Gemini-South Telescope. FLAMINGOS-2 has a pixel scale of $0.18''$ pixel$^{-1}$ and offers imaging in the $0.9-2.5\mu$m wavelength range within a $6.1'$ circular field of view. These observations were taken between November 2023 and January 2024 with more than ten epochs in each $JHK_s$ filter. These time-series observations in queue mode typically consisted of 11 and 22 dithered frames\footnote{In some cases, more than the requested number of dithered were obtained because of variation in the observing conditions during the sequence. Table~\ref{tbl:data} lists final number of {\it Usable} frames at a given epoch.} at each location in $J$ and $H/K_s$ bands, respectively. Each dithered exposure was 10s in all three filters and the typical time of the entire dithered sequence was $~\sim 8/12$ minutes in $J/HK_s$ including overheads. We utilized a total of 141 science frames in $J$, 242 in $H$, and 231 in $K_s$ for this analysis. A log of $JHK_s$ observations is provided in Table ~\ref{tbl:data}.

The raw science images and associated calibration frames (Darks and Flats) were downloaded from the Gemini Observatory Archive{\footnote{\url{https://archive.gemini.edu/}}}.
We used the DRAGONS{\footnote{\url{https://zenodo.org/records/17412281}}} (Data Reduction for Astronomy from Gemini Observatory North and South) pipeline \citep{labrie2023, dragons1025} to pre-process the images including bad-pixel masking, non-linearity correction, dark subtraction, and flat-fielding. The sky background images were constructed using DRAGONS and were further cleaned for residual variations using the \texttt{SExtractor} \citep{bertin1996}. Given the faintness of our target RRL in Reticulum, all dithered frames at a given epoch were stacked to create a science ready epoch image in each filter. A reference median combined image with astrometric calibration was also created from best-seeing epoch images. The details of point-spread function (PSF) photometry using \texttt{DAOPHOT/ALLSTAR} \citep{stetson1987} and \texttt{ALLFRAME} \citep{stetson1994} routines have been provided in \citet{bhardwaj2022a} based on Gemini-F2 data and will not be repeated here. The only difference is that in \citet{bhardwaj2022a}, PSF photometry was performed on individual dithered frames, while in this work we stacked images at each epoch due to faintness of RRL in Reticulum. Similarly, the photometric calibration was performed using 2MASS catalog in $JHK_s$ bands as discussed in detail in \citet{bhardwaj2022a} and \citet{Bhardwaj2024a}.

\subsection{Optical-NIR color--magnitude diagrams}

Deep NIR photometry of Reticulum was combined with the HST photometry in F606W and F814W filters from \citet{milone2023} covering central $\sim3.4\arcmin\times3.4\arcmin$ with the Wide Field Chanel of Advanced Camera for Surveys. While only 1776 sources were detected in Gemini-F2 reference image, there are more than 12000 sources with both F606W and F814W photometry within the common field of view. The combined optical-NIR color--magnitude diagrams for 1156 sources that match within $1\arcsec$ tolerance, as well as for HST only sources are shown in Fig.~\ref{fig:cmd}. In the magnitude range covering horizontal branch, the median uncertainties are 0.01 and 0.02 mag in $J$ and $K_s$ bands, respectively. This is reflected in a well-defined horizontal branch slope in the NIR and a narrow distribution of stars along the red giant branch. We note that the {\it HST} photometry is based on random-epoch observations which will reflect in the scatter in the optical magnitudes and colors for variable horizontal branch stars.

The magnitudes in the color--magnitude diagram were corrected for extinction assuming the reddening law of \citet{card1989} and an $R_V=3.23$. This results in the total-to-selective absorption ratios in NIR filters, $A_{J/H/K_s} = 0.94/0.58/0.39E(B - V)$, as utilized in the previous works on NIR photometry and RRL period-luminosity-metallicity calibrations \citep[e.g.,][]{bhardwaj2023}. The color-excess of $E(B-V)=0.03$ mag was adopted following \citet{muraveva2018b}. This value was determined by \citet{walker1991} using the minimum color of RRL stars and is close to the average value in the literature \citep{muraveva2018b}. Note that the extinction in the NIR is systematically small and a variation of $\pm0.03$~mag in color-excess amounts to $\sim0.01$ mag offset in the $K_s$ band. We also used the total-to-selective absorption ratios of 2.47 and 1.53 in the HST F606W and F814W filters, respectively \citep{schlafly2011}. 

Fig.~\ref{fig:cmd} shows zero-age horizontal branch models from BaSTI{\footnote{\url{http://basti-iac.oa-abruzzo.inaf.it/index.html}}} (Bag of Stellar Tracks and Isochrones) for a mass range of $0.5-0.8M_\odot$ using models with $\alpha$-enhanced composition \citep{pietrinferni2021}. These models cover a wide range of metallicities from [Fe/H]$\sim-1.40$ to $-1.90$~dex and a helium content of $Y=0.245$. The photometrically transformed bolometric magnitudes to the 2MASS and HST filters were offset with the precise geometric distance to the LMC \citep[$18.477\pm0.026$~mag,][]{piet2019}, and show a good agreement with the observed horizontal branch. An increase in the metallicity from $-1.90$ to $-1.40$~dex leads to a systematic decrease in the mean-luminosity level or magnitudes on the horizontal branch and the RRL instability strip \citep{marconi2015}. There are 24 RRL stars from \citet{kuehn2013} in the common optical-NIR sample that are also shown in the bottom panels. The observed scatter around the horizontal branch tracks is due to the variable nature and random-epoch optical data for RRL stars.

\begin{figure}
\centering
\includegraphics[width=0.48\textwidth]{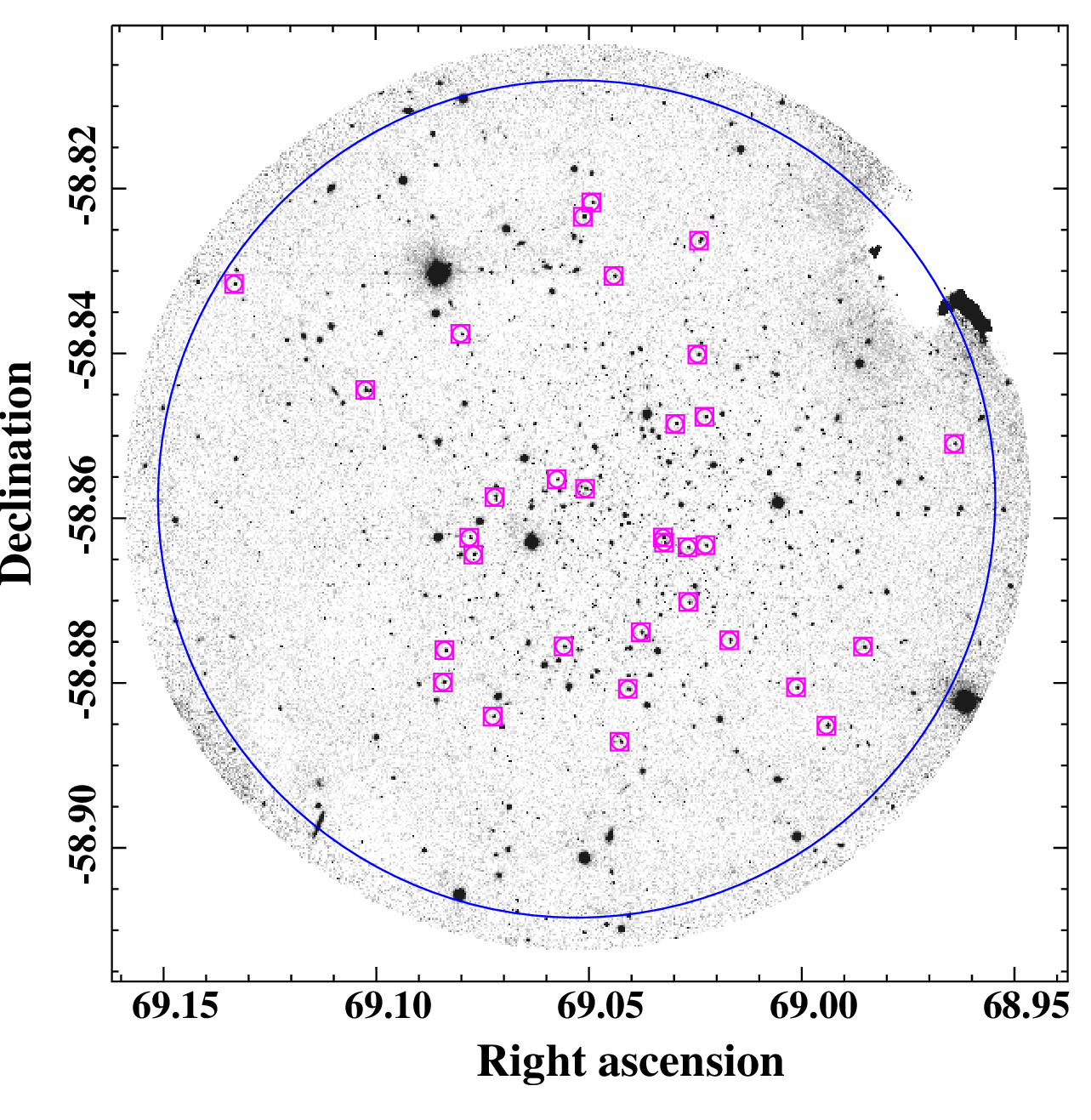}
\caption{Spatial distribution of RRL (in magenta squares) in Reticulum in the reference image. The blue circle represents $6.1\arcmin$ circular field of view of Gemini-F2.}
\label{fig:spatial}
\end{figure}

\begin{figure*}[ht!]
\centering
 \includegraphics[width=0.99\textwidth]{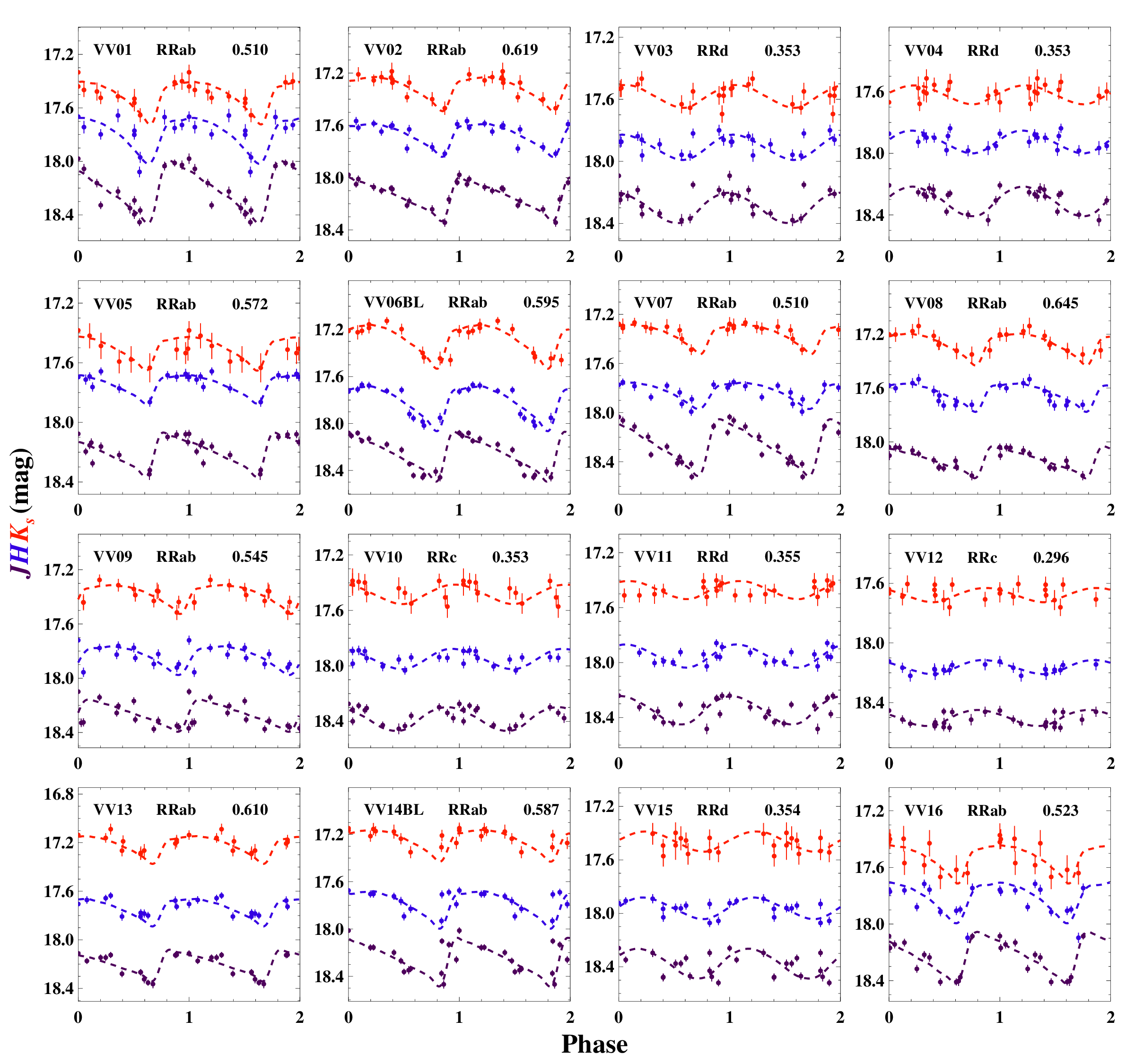}
 \caption{Phased light curves of RRL stars in Reticulum in the NIR bands. The $H$-band (blue) and $K_s$-band (red) light curves are offset for clarity by $-0.2$ mag and $-0.6$~mag, respectively. The dashed lines represent the best-fitting templates to the light curve data in each band. The star ID, variable subtype, and the pulsation period are included at the top of each panel.}
\label{fig:rrl_lc}
\end{figure*}

\begin{figure*}[ht!]
\centering
 \includegraphics[width=0.99\textwidth]{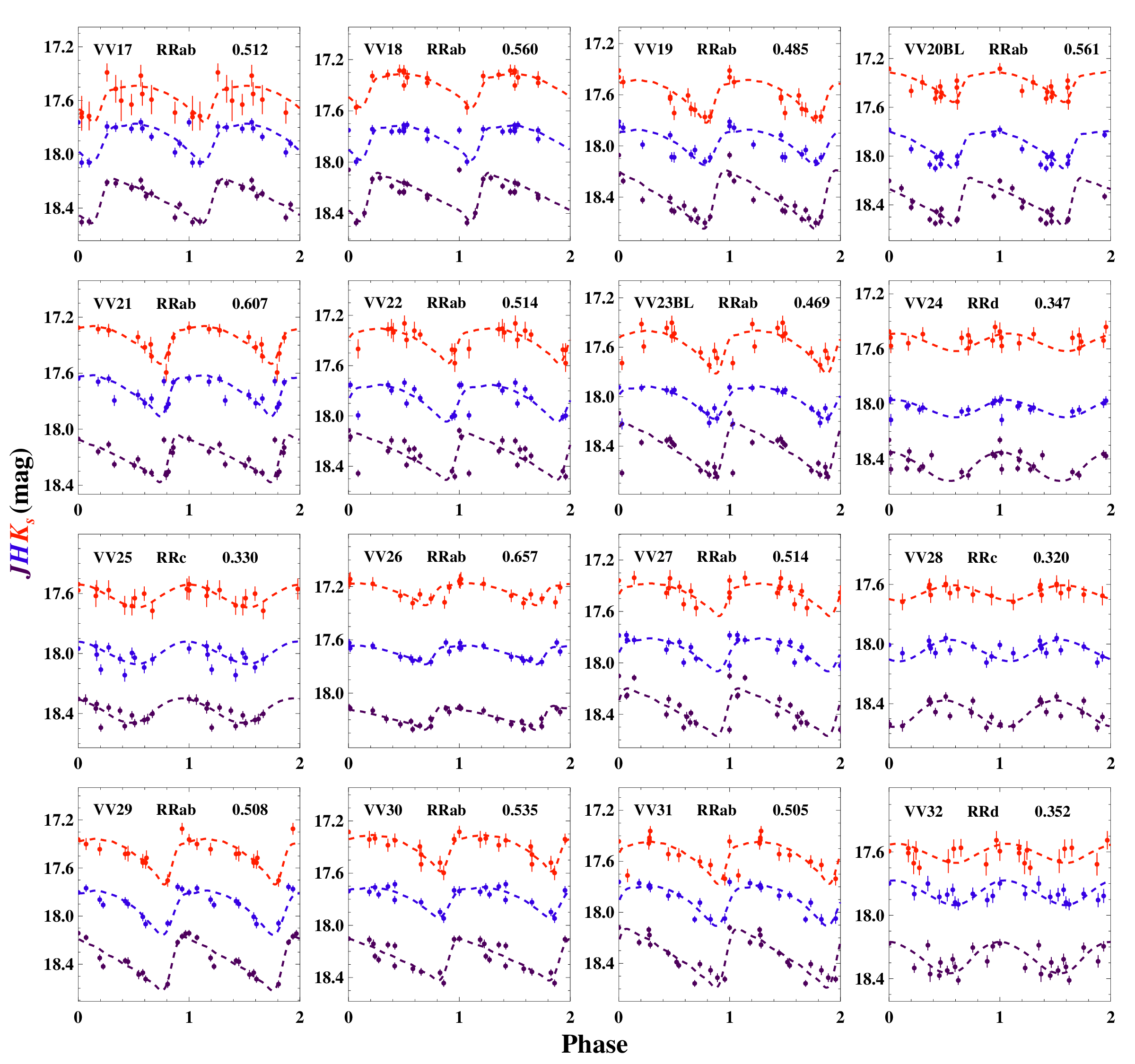}
\caption{Same as in Fig.~\ref{fig:rrl_lc}.}.
\label{fig:rrl_lc1}
\end{figure*}

We also cross-matched the sources with NIR photometry to the Gaia DR3 catalog \citep{vallenari2022} and found 316 common stars within a tolerance of $1\arcsec$. The mean proper motions along the right ascension and declination axes were found to be $1.992\pm0.020$ and $-0.317\pm0.023$ mas/yr with a dispersion of 0.35 and 0.39 mas/yr, respectively. These mean proper motions of Reticulum cluster are in good agreement with independent proper motions based on {\it HST} data \citep[e.g., $1.964\pm0.093$ and $-0.307\pm0.099$ mas/yr in][]{benett2022}. Assuming that all sources within the $3\sigma$ scatter of the mean proper motions are likely members, we find only 207 cluster stars in optical-NIR sample with Gaia. This small sample is not shown in Fig.~\ref{fig:cmd}, but since Gaia magnitudes are based on multi-epoch observations, less spread is seen in the expected variable horizontal branch (see section~\ref{sec:var_rrl}). 

\section{RR Lyrae stars in Reticulum}
\label{sec:var_rrl}

\citet{kuehn2013} published optical photometry of 32 RRL variables stars (22 RRab, 4 RRc, and 6 RRd) in the Reticulum cluster along with their positions and pulsation properties. This initial list of 32 variables was cross-matched with our NIR photometric catalogue within $1\arcsec$ tolerance. All of the RRL variables were found within the Gemini-F2 field of view and their spatial distribution is shown in Fig.~\ref{fig:spatial}. The $JHK_s$ light curves for each RRL were extracted from time-series photometry. The light curves were phased using the periods adopted from \citet{kuehn2013} with a reference epoch of $J$-band maximum at phase zero. The RRd stars were phased with their dominant first-overtone periods. 

The phased light curves of RRL were fitted with templates in the $JHK_s$ bands taken from \citet{braga2019}. The template light curves for RRab and RRc stars in the $JHK_s$ were fitted with variable amplitude and phase parameters solving for the mean-magnitudes, amplitude, and a phase offset with respect to maximum light. The phase-offset for $HK_s$ band template fit was fixed with the value obtained in the $J$-band. Figs~\ref{fig:rrl_lc}-\ref{fig:rrl_lc1} display template-fitted  $JHK_s$ light curves of all RRL variables in Reticulum. 

The light curves are well-sampled with RRab stars clearly exhibiting saw-tooth light curves, while RRc/RRd stars show nearly sinusoidal brightness variations. The RRL light curves exhibit larger scatter given their fainter magnitudes in the $K_s$ band. Table~\ref{tbl:var} provides the pulsation properties and intensity-averaged magnitudes as well as amplitudes for RRL stars. The mean magnitudes in the $JHK_s$ bands for RRL in Reticulum have the average uncertainty of 0.035 mag.

\begin{figure}
\centering
  \includegraphics[width=0.48\textwidth]{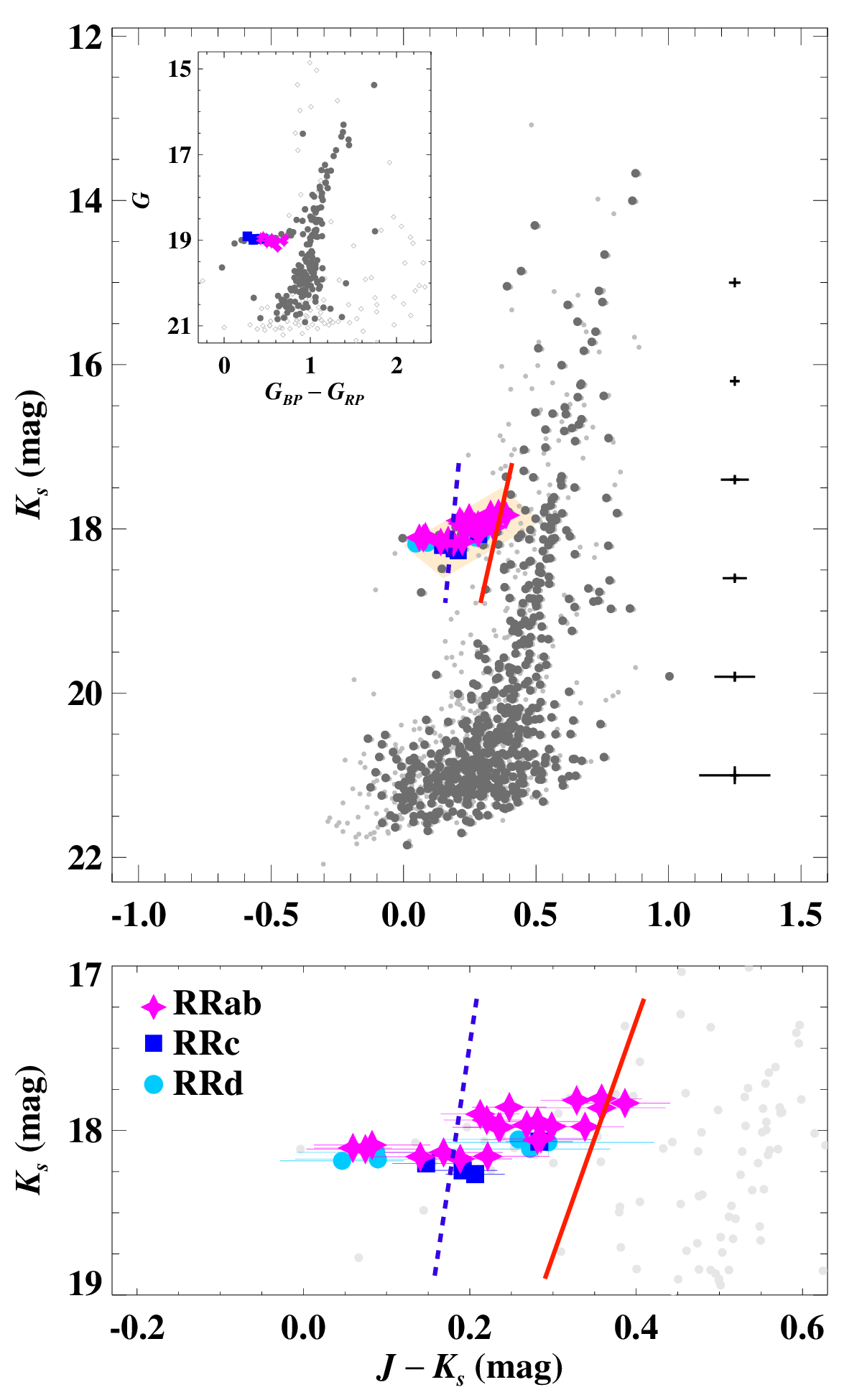}
  \caption{Extinction-corrected $(J-K_s),K_s$ color--magnitude diagram for Reticulum cluster (small grey dots). The larger circles display likely cluster members as discussed in the text. RRL stars are also overplotted with their mean magnitudes and colors. The solid and dashed lines display theoretically predicted instability strip boundaries - fundamental red edge and first-overtone blue edge, respectively \citep{marconi2015}. The representative $\pm3\sigma$ error bars are shown as a function of magnitude and color. The color--magnitude diagram in the {\it Gaia} bands for common sources is also shown in the inset figure. Larger symbols represent cluster member based on proper motion selection. Bottom panel displays zoomed-in view of the horizontal branch and RRL stars.}
  \label{fig_cmd}
\end{figure}

\begin{figure}
\centering
  \includegraphics[width=0.47\textwidth]{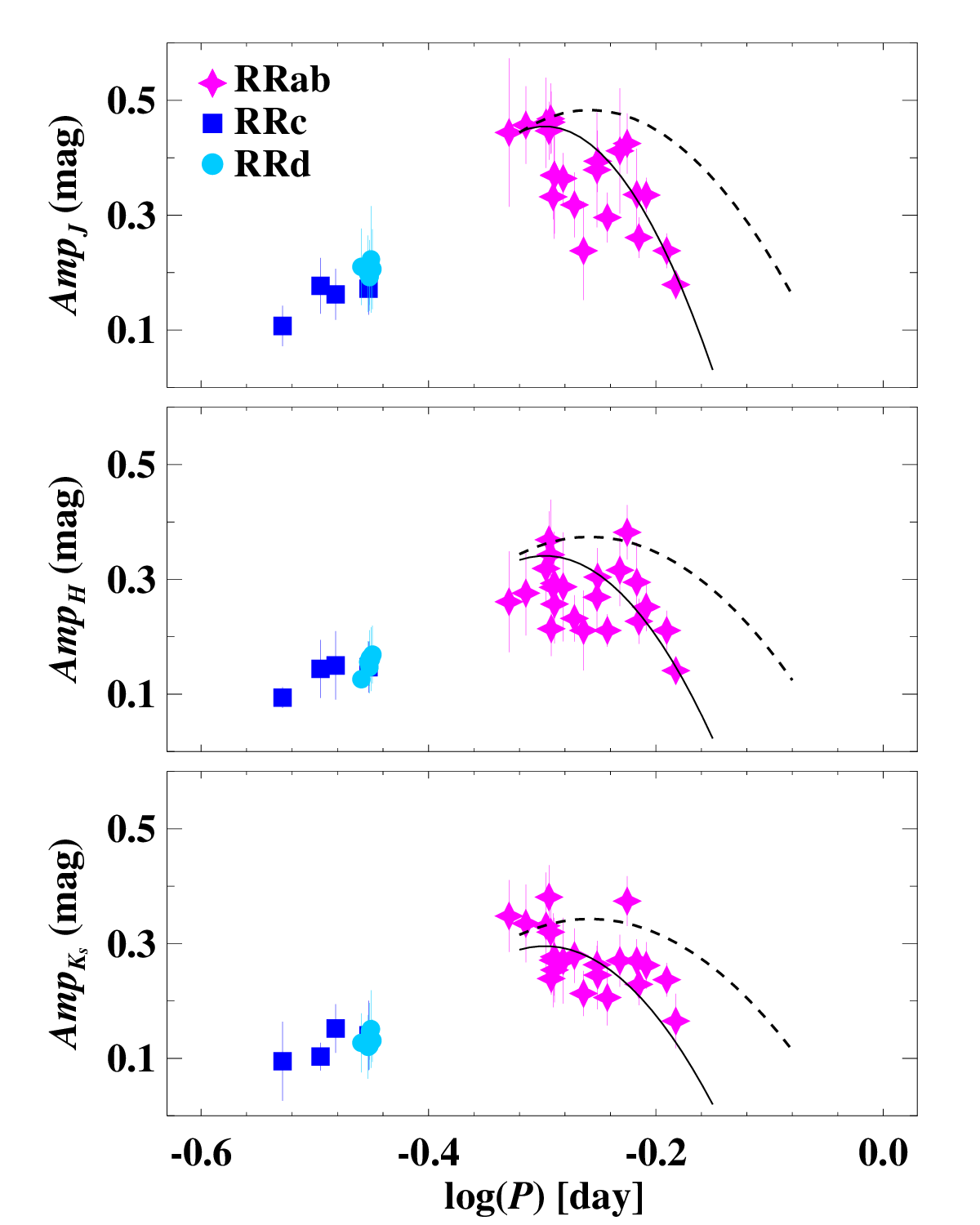}
  \caption{Bailey diagrams for RRL stars in Reticulum in the $J$ (top), $H$ (middle), and $K_s$ (bottom) bands. The solid and dashed lines represent approximate $JHK_s$ loci of OoI and OoII clusters from \citet[][M3]{bhardwaj2020} and \citet[][M53]{bhardwaj2021}, respectively. The uncertainties on the amplitudes represent $1\sigma$ scatter around the best-fitting template light curve.}
  \label{fig_bailey}
\end{figure}

\begin{deluxetable*}{lccclccccccccc}
\tablecaption{NIR pulsation properties of RRL variable stars in Reticulum. \label{tbl:var}}
\tabletypesize{\small}
\tablewidth{0pt}
\tablehead{
{ID} & {RA} & {Dec} & {P} &  {Type}& \multicolumn{3}{c}{Mean magnitudes ($m_\lambda$)}  & \multicolumn{3}{c}{$\sigma_{m_\lambda}$}& \multicolumn{3}{c}{Amplitudes ($m_\lambda$)}  \\
 	&	&   &    &	   & $J$  &   $H$  & $K_s$  & $J$  &  $H$  & $K_s$ & $J$  &  $H$  & $K_s$ 	\\  
 	&	deg.&	deg.	& days 	    &	   & \multicolumn{3}{c}{mag}  & \multicolumn{3}{c}{mag}  & \multicolumn{3}{c}{mag}}
  \startdata
     VV01&    68.964583&   -58.850944&    0.50993&      RRab&   18.199&   17.989&   18.100&    0.061&    0.035&    0.034&    0.468&    0.343&    0.320\\
     VV02&    68.985833&   -58.875583&    0.61869&      RRab&   18.141&   17.870&   17.912&    0.035&    0.031&    0.032&    0.335&    0.252&    0.262\\
     VV03&    68.994167&   -58.885194&    0.35350&       RRd&   18.248&   18.082&   18.144&    0.075&    0.058&    0.059&    0.195&    0.163&    0.147\\
     VV04&    69.001250&   -58.880556&    0.35320&       RRd&   18.259&   18.117&   18.196&    0.054&    0.055&    0.051&    0.192&    0.147&    0.122\\
     VV05&    69.017083&   -58.874806&    0.57185&      RRab&   18.194&   17.945&   18.118&    0.032&    0.024&    0.034&    0.296&    0.211&    0.206\\
   VV06BL&    69.022500&   -58.863278&    0.59526&      RRab&   18.249&   18.004&   17.874&    0.059&    0.058&    0.049&    0.425&    0.382&    0.374\\
     VV07&    69.022917&   -58.847722&    0.51044&      RRab&   18.255&   18.029&   17.957&    0.057&    0.032&    0.024&    0.462&    0.214&    0.239\\
     VV08&    69.023750&   -58.826194&    0.64496&      RRab&   18.134&   17.834&   17.870&    0.031&    0.033&    0.032&    0.238&    0.211&    0.237\\
     VV09&    69.024583&   -58.840139&    0.54496&      RRab&   18.263&   18.037&   17.978&    0.035&    0.032&    0.031&    0.238&    0.211&    0.213\\
     VV10&    69.026667&   -58.870167&    0.35256&       RRc&   18.379&   18.148&   18.079&    0.028&    0.024&    0.029&    0.172&    0.147&    0.140\\
     VV11&    69.026667&   -58.863528&    0.35540&       RRd&   18.341&   18.152&   18.066&    0.080&    0.059&    0.047&    0.206&    0.169&    0.131\\
     VV12&    69.029583&   -58.848583&    0.29627&       RRc&   18.500&   18.360&   18.277&    0.024&    0.022&    0.026&    0.107&    0.094&    0.095\\
     VV13&    69.032500&   -58.863000&    0.60958&      RRab&   18.193&   17.937&   17.818&    0.039&    0.032&    0.029&    0.261&    0.227&    0.229\\
   VV14BL&    69.032500&   -58.862333&    0.58661&      RRab&   18.248&   17.982&   17.845&    0.046&    0.034&    0.029&    0.412&    0.316&    0.270\\
     VV15&    69.037917&   -58.873833&    0.35430&       RRd&   18.395&   18.168&   18.084&    0.102&    0.064&    0.076&    0.223&    0.161&    0.151\\
     VV16&    69.040833&   -58.880750&    0.52290&      RRab&   18.216&   18.004&   18.125&    0.052&    0.041&    0.048&    0.364&    0.287&    0.270\\
     VV17&    69.042917&   -58.887111&    0.51241&      RRab&   18.328&   18.068&   18.171&    0.040&    0.041&    0.041&    0.332&    0.286&    0.271\\
     VV18&    69.044167&   -58.830611&    0.56005&      RRab&   18.243&   18.000&   17.992&    0.034&    0.023&    0.025&    0.379&    0.269&    0.263\\
     VV19&    69.049583&   -58.821694&    0.48485&      RRab&   18.389&   18.166&   18.184&    0.042&    0.046&    0.046&    0.457&    0.276&    0.335\\
   VV20BL&    69.050833&   -58.856472&    0.56075&      RRab&   18.344&   18.094&   17.989&    0.037&    0.031&    0.029&    0.394&    0.304&    0.245\\
     VV21&    69.051250&   -58.823417&    0.60700&      RRab&   18.185&   17.909&   17.948&    0.037&    0.038&    0.036&    0.336&    0.295&    0.270\\
     VV22&    69.055833&   -58.875556&    0.51359&      RRab&   18.302&   18.051&   17.987&    0.041&    0.042&    0.039&    0.370&    0.293&    0.277\\
   VV23BL&    69.057500&   -58.855278&    0.46863&      RRab&   18.407&   18.206&   18.169&    0.056&    0.049&    0.049&    0.444&    0.261&    0.348\\
     VV24&    69.072083&   -58.857389&    0.34750&       RRd&   18.413&   18.211&   18.124&    0.076&    0.029&    0.059&    0.210&    0.126&    0.127\\
     VV25&    69.072500&   -58.884056&    0.32991&       RRc&   18.376&   18.193&   18.212&    0.029&    0.027&    0.028&    0.162&    0.150&    0.152\\
     VV26&    69.077083&   -58.864389&    0.65696&      RRab&   18.173&   17.889&   17.828&    0.029&    0.028&    0.032&    0.179&    0.141&    0.165\\
     VV27&    69.077917&   -58.862306&    0.51382&      RRab&   18.361&   18.096&   18.059&    0.046&    0.033&    0.031&    0.369&    0.257&    0.254\\
     VV28&    69.080000&   -58.837639&    0.31994&       RRc&   18.462&   18.241&   18.254&    0.033&    0.025&    0.025&    0.177&    0.144&    0.103\\
     VV29&    69.083750&   -58.875972&    0.50815&      RRab&   18.367&   18.110&   18.069&    0.055&    0.041&    0.037&    0.447&    0.369&    0.381\\
     VV30&    69.084167&   -58.879917&    0.53501&      RRab&   18.245&   17.986&   17.992&    0.043&    0.031&    0.038&    0.318&    0.232&    0.279\\
     VV31&    69.101667&   -58.844472&    0.50516&      RRab&   18.330&   18.094&   18.145&    0.053&    0.042&    0.047&    0.461&    0.319&    0.331\\
     VV32&    69.133333&   -58.831444&    0.35230&       RRd&   18.293&   18.069&   18.187&    0.077&    0.045&    0.063&    0.198&    0.156&    0.120\\
\enddata
\tablecomments{'BL' in Star ID refers to Blazhko effect in RRL variables. The period and subtypes are adopted from \citet{kuehn2013}.\\}
\end{deluxetable*}

\begin{figure*}
\centering
\includegraphics[width=0.98\textwidth]{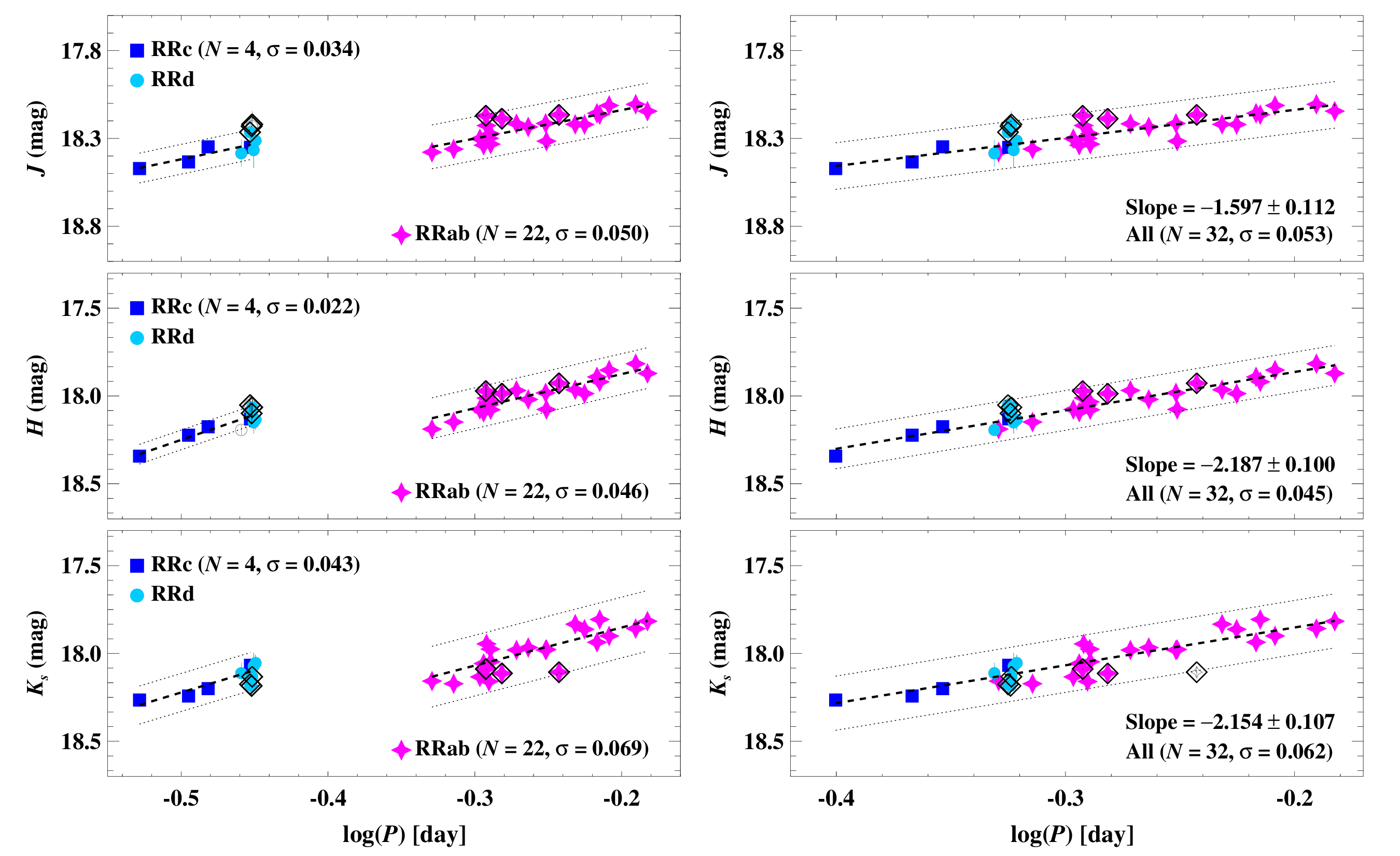}
	\caption{$JHK_s$ period-luminosity relations for RRab and RRc stars (left) and all RRL stars (right) in $J$ (top), $H$ (middle), and $K_s$ (bottom) in Reticulum. In the right panels, the periods for the RRc/RRd stars have been shifted to their corresponding fundamental-mode periods, as explained in the text. The dashed lines show best-fitting linear regressions over the period range under consideration, and the parallel dotted lines display $\pm 2.5\sigma$ offsets. Six RRL that are bluer than $J-K_s <0.1$~mag in Fig~\ref{fig_cmd}, are shown using open diamonds.\\} 
\label{fig:plr_rrl}
\end{figure*}

\subsection{RR Lyrae and the horizontal branch topology}

The intensity-averaged magnitudes for RRL stars were corrected for extinction in the NIR color--magnitude diagram for Reticulum cluster shown in Fig.~\ref{fig_cmd}. Note that not all of the NIR sources are available in optical catalogs as discussed in Section~\ref{sec:data}. Therefore, we used color-color diagram to remove likely field stars and possible spurious point sources. We obtained extinction-corrected median colors of $J-H = 0.24$ mag and $H-K_s = 0.11$ mag with a scatter of $\sigma=0.12$ mag. All stars located within the $5\sigma$ circular radius of the color-color centroid are considered cluster members and are shown as larger gray symbols in Fig.~\ref{fig_cmd}. For a relative comparison, proper-motion cleaned Gaia color--magnitude diagram is also shown using the common sources discussed in Section~\ref{sec:data}. The horizontal branch is well-populated and narrow distribution along red giant branch is clearly visible in both optical and NIR color--magnitude diagrams.

The spatial distribution of all 32 RRL stars in the instability strip in the color--magnitude diagram is also shown. The predicted red and blue edges of the instability strip were derived using theoretical RRL models in \citet{marconi2015}. We apply a distance modulus offset corresponding to the geometric distance to the LMC from \citet{piet2019}. The bottom panel shows that most of Reticulum RRL are located on the horizontal branch and are within the predicted fundamental red edge and the first-overtone blue edge, given the observational errors in the mean magnitudes and colors. Six RRL variables (3 RRab and 3 RRd) appear systematically bluer ($J-K_s < 0.1$~mag) -- these stars are located in the portion of the Gemini-F2 images that have increased background in $K_s$-band, leading to increased sky subtraction residuals and larger uncertainty in photometric measurements in $K_s$-band\footnote{In the $K_s$-band, a small portion in the Gemini-F2 images show a bright bubble that arises due to the scattering light pattern. This has been verified with the DRAGON helpdesk and stars located in this bubble are likely affected by additional sky subtraction uncertainties.}. Their impact on PL analysis is discussed in Section~\ref{sec:plr_rrl}.

\subsection{NIR Bailey diagrams and Oosterhoff classification}

Reticulum was classified as Oo-I type cluster by \cite{kuehn2013}, based on the mean period of its RRab stars and its optical period-amplitude diagrams, also known as Bailey diagrams \citep{bailey1902} . We show here for the first time the NIR Bailey diagrams for Reticulum in Fig.~\ref{fig_bailey}. Most of RRab follow the locus of period-amplitude diagram for M3 ([Fe/H]$\sim-1.5$~dex) from \citet{bhardwaj2020a}, confirming the expected trend for OoI type clusters. Approximate locus of OoII type M53 ([Fe/H]$\sim-2.06$~dex) cluster from \citet{bhardwaj2021} are also shown for comparison. The uncertainties on the amplitude measurements represent $1\sigma$ scatter around the best-fitting template light curve.

\subsection{Period-Luminosity relations}
\label{sec:plr_rrl}

The intensity-averaged magnitudes for RRL were used to derive PL relations in three separate subgroups: (1) RRab only, (2) RRc only, and (3) a combined sample of RRab+RRc+RRd. The dominant first-overtone mode periods were used for RRd stars for the combined sample analysis. For the fundamentalization, we use the equation, $\log(P_{{RRab}})=\log(P_{{RRc/RRd}})+0.127$  \citep{iben1974, braga2022}. 

We fit a PL relation under the assumption of linearity over the entire period range. NIR $JHK_s$ PL relations for RRL stars in Reticulum are shown in Fig.~\ref{fig:plr_rrl}. The residuals of all RRL were found to be within $3\sigma$ scatter of the PL relation. The scatter in these relations is of the order of $0.05$~mag and primarily reflects the intrinsic width of the instability strip as well as contribution from photometric uncertainties on mean-magnitudes (0.035 mag). The Galactic GCs studied in \citet{bhardwaj2023} also exhibit similar scatter and further suggest that the metallicity spread is small or negligible in Reticulum cluster. The coefficients of the PL relations are listed in Table~\ref{tbl:plr_rrl}.

\begin{deluxetable}{cccccc}
\tablecaption{Near-infrared PL relations of RRL in Reticulum. \label{tbl:plr_rrl}}
\tablewidth{0pt}
\tablehead{
{~~Band~~} & {Type}& {~~$\alpha_\lambda$~~} & {~~$\beta_\lambda$~~} & {~~$\sigma$~~}& {~~$N$}~~}
\startdata
     $J$ &  RRab &    17.814$\pm$0.053      &$     -1.621\pm0.210      $&      0.050 &   22\\
     $J$ &   RRc &    17.540$\pm$0.236      &$     -1.757\pm0.477      $&      0.034 &    4\\
     $J$ &   All &    17.818$\pm$0.033      & $    -1.597\pm0.112      $&      0.053 &   32\\
     $H$ &  RRab &    17.486$\pm$0.048      &$     -1.946\pm0.189      $&      0.046 &   22\\
     $H$ &   RRc &    16.815$\pm$0.209      &$     -2.871\pm0.424      $&      0.022 &    4\\
     $H$ &   All &    17.426$\pm$0.029      &$     -2.187\pm0.100      $&      0.045 &   32\\
   $K_s$ &  RRab &    17.419$\pm$0.048      &$     -2.168\pm0.188      $&      0.069 &   22\\
   $K_s$ &   RRc &    16.963$\pm$0.248      &$     -2.518\pm0.504      $&      0.043 &    4\\
   $K_s$ &   All &    17.421$\pm$0.031      &$     -2.154\pm0.107      $&      0.062 &   32\\
\enddata
\tablecomments{The PL relation were fitted in the following form: $m_\lambda = a_\lambda + b_\lambda \log(P)$, where $m_\lambda$ is the mean magnitude in $JHK_s$ bands, and $a_\lambda$ and $b_\lambda$ give the zero-point and the slope of the PL relation. The dispersion ($\sigma$) and the number of stars ($N_{i/f}$) in the initial/final PL fits are also tabulated.\\}
\end{deluxetable}

The $J$-band PL relation for Reticulum RRL has a significantly shallower slope when compared to that of Galactic GCs. For example, the combined slope of the M3 cluster RRL is $-1.830\pm0.031$ \citep{bhardwaj2020a}. We also note that Reticulum PL slopes are systematically shallower (and show a larger scatter due to lower statistics) in comparison with the global slope of the RRL PLZ relations derived in 11 Galactic GCs \citep[$-1.83\pm0.02$ in $J$, $-2.29\pm0.02$ in $H$, and $-2.37\pm0.02$ in $K_s$,][]{bhardwaj2023}. Similarly, Reticulum PL slopes listed in Table~\ref{tbl:plr_rrl} are shallower than the slopes of RRL PL relations in Draco and Carina dwarf spheroidal galaxies \citep{Bhardwaj2024a, ngeow2026}. The slope of $K_s$-band PL relation is in excellent agreement with the previous study of \citet[$-2.15\pm0.09$ mag/dex,][]{dallora2004}, but the apparent zero-point is $\sim1.5\sigma$ fainter for both RRab and the combined sample of RRL stars. 
Note that we use the full sample of 32 RRL, while \citet{dallora2004} used 30 stars for their study. The zero-point difference is reduced if the RRL with the largest residual (VV05, see Fig.~\ref{fig:plr_rrl}) or the RRc with the shortest period (VV12) are excluded from the sample. Fig.~\ref{fig:plr_rrl} also shows six RRL stars (open diamonds) that exhibit systematically bluer colors ($J-K_s < 0.1$~mag) than the instability strip boundaries in Fig.~\ref{fig_cmd}. It is evident that these stars are brighter than best-fit PL relations in the $JH$ bands but become fainter than the best-fit relation in the $K_s$-band. The RRL (VV05) with the largest residual in $K_s$ is also among these six stars. If we exclude all of these RRL stars from the PL analysis, the zero-point becomes brighter by $0.6\sigma$ for the combined sample, but no changes are seen in the zero-point for RRab sample. Therefore, the difference in zero-point with \citet{dallora2004} remain at $\sim1.5\sigma$ for RRab sample and reduces to $1.2\sigma$ for the combined sample.

\section{Distance to the Reticulum Cluster}
\label{sec:dis_retic}

\begin{figure}
\centering
  \includegraphics[width=0.48\textwidth]{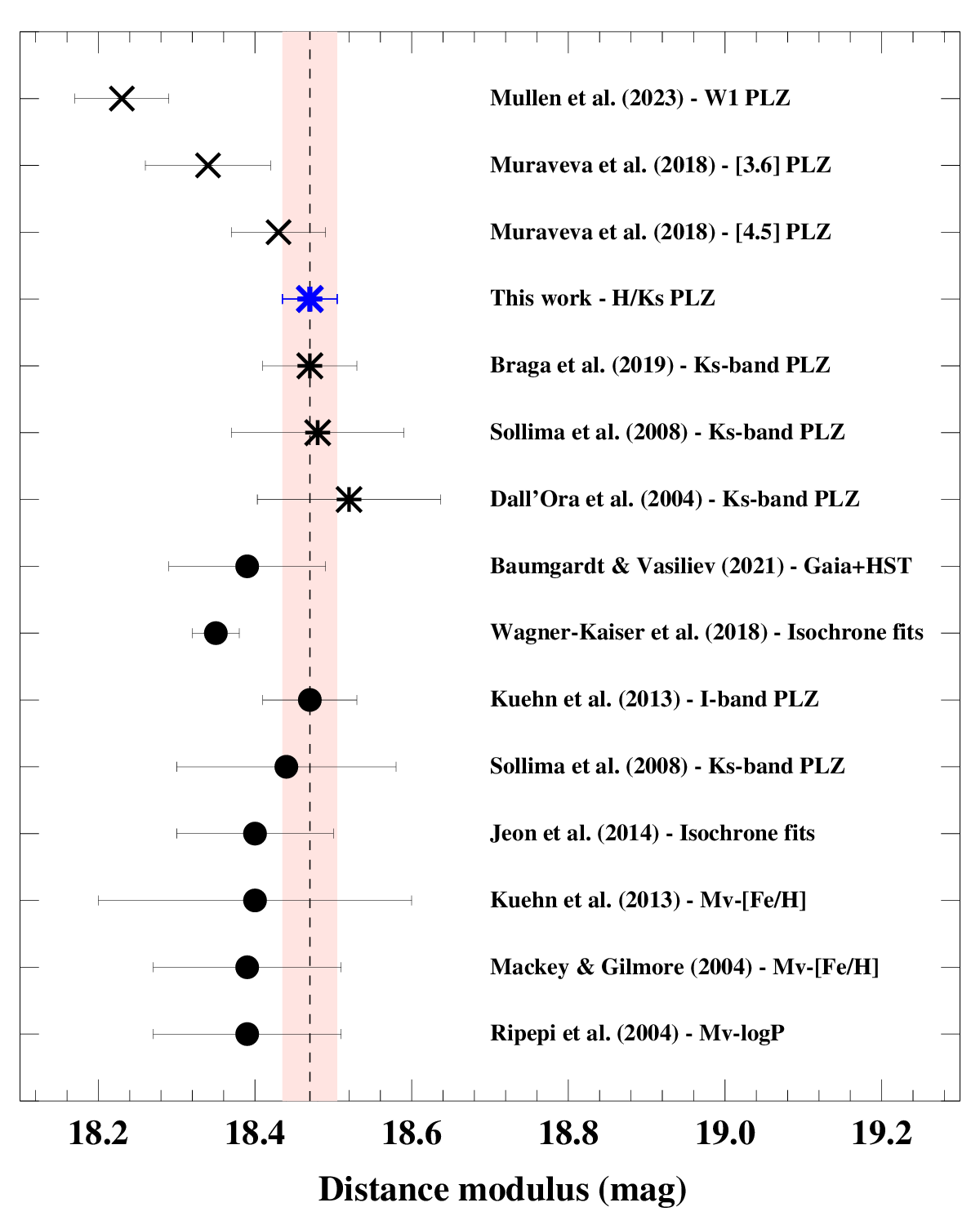}
  \caption{Distance moduli to Reticulum cluster available in the literature based on optical (circles), NIR (asterisks), and MIR (crosses) data: \citep{ripepi2004, mackey2004, kuehn2013, jeon2014, sollima2008, wagner2017a, baumgardt2021, dallora2004, braga2019, muraveva2018b, mullen2023}. The dashed line and shaded region represent the distance modulus and associated error from this work.}
  \label{fig_mu}
\end{figure}

\subsection{Literature distance and metallicity measurements}

Modern distance measurements to the Reticulum cluster based on multiwavelength datasets and a variety of independent methods are shown in Fig.~\ref{fig_mu}. When compared with the precise late-type eclipsing binary distance to the LMC of $18.477\pm0.026$~mag, most distance modulus estimates locate Reticulum $\sim1-3$~kpc in front of the LMC barycenter. The distance determination of $18.52\pm0.12$~mag based on $K_s$-band photometry of RRL in \citet{dallora2004} is very relevant to this work, and it is the only distance placing Reticulum beyond the LMC center. Using the same $K_s$-band data and improved empirical calibrations of RRL PLZ relations and template-fits, \citet{sollima2008} and \citet{braga2019} obtained a distance to Reticulum that is closer to the LMC barycenter distance. However, \citet{muraveva2018b} used mid-infrared observations of RRL from the Spitzer Space Telescope anchored to parallax calibrations, and found that Reticulum lies approximately 3 kpc in the foreground of the LMC barycenter. Using similar empirical MIR PL relations of RRL stars calibrated using Gaia DR3 parallaxes, \citet{mullen2023} found a distance to Reticulum that is significantly smaller than the previous literature measurements taken from a variety of methods. Most distance estimates based on optical, NIR, and mid-infrared photometry are statistically consistent given the relatively large uncertainties that are intrinsic to different calibrations, anchors, and approaches adopted in literature studies.

Most literature studies on distance measurements adopted metallicity values to the Reticulum cluster that range between $-1.5$ and $-1.7$~dex in different metallicity scales. \citet{suntzeff1992} used spectroscopic Ca II triplet metallicities of individual stars in Reticulum and obtained a [Fe/H]$_{ZW}\sim-1.71$~dex, this was the value used in \citet{dallora2004} for distance determination. \citet{mackey2004} used HST color--magnitude diagrams to estimate a metallicity of 
$-1.66\pm0.12$~dex while \citet{kuehn2013} used Fourier analysis of RRab stars obtaining a value between $-1.5$ and $-1.7$~dex with different empirical relations. More recently, \citet{sarajedini2024} obtained metallicities of old LMC clusters based on period-amplitude metallicity relation and found a value of $-1.55\pm0.03$ to Reticulum in the \citet{zinn1984} scale. For a relative comparison with \citet{dallora2004}, we also adopt the mean-metallicity of $-1.71\pm0.10$~dex to the cluster. However, the empirical calibration of PLZ relations in \citet{bhardwaj2023} use the metallicity scale of \citet{carretta2009}. Using the transformations from \citet{carretta2009}, we obtain a [Fe/H]$_{UVES}\sim-1.67\pm0.12$~dex for the Reticulum cluster.

\subsection{RR Lyrae based distance}

We utilized our homogeneous NIR observations of RRL stars to derive an accurate and precise distance modulus for the Reticulum cluster. The calibrated PLZ relations for RRL were taken from \citet{bhardwaj2023}, who provided the most precise empirical determination of the metallicity dependence of $JHK_s$ band PL relations using 964 RRL in 11 Galactic GCs. These empirical calibrations were not only in excellent agreement with predictions of the horizontal branch models  \citep{catelan2004} as well as stellar pulsation models \citep{marconi2015}, but also offer a unique advantage that some of the calibration sample and Reticulum observations were obtained using the same instrument, Flamingos-2 on the Gemini South Telescope, and were reduced following a consistent methodology. This homogeneous observational framework minimizes potential systematic offsets arising from instrumental effects, photometric zero points, and data reduction procedures, thus enabling a robust and precise cluster distance determination.

The extinction-corrected mean magnitudes together with the adopted mean-metallicity were used to derive absolute magnitudes in $JHK_s$ bands using calibrator PLZ relations from \citet{bhardwaj2023}. Since there are only four RRc stars in Reticulum, we used the combined sample of RRL to derive distances. We obtained distance moduli of $18.53\pm0.04$, $18.45\pm0.03$, and $18.46\pm0.03$ in $JHK_s$ bands, respectively. We also use the theoretical calibrations in the $JHK_s$-bands from \citet{marconi2015} to derive distance moduli of $18.53\pm0.03$, $18.48\pm0.030$, and $18.51\pm0.03$, respectively. Previous studies of Galactic GCs and dwarf spheroidal galaxies have shown that the theoretical and empirical calibration based on $J$-band PLZ relation systematically provide $1\sigma$ larger distance modulus \citep{Bhardwaj2024a, ngeow2026}. This is also reflected in a difference of $0.06$~mag between absolute zero-points in the $J$-band from \citet{bhardwaj2023} and \citet{marconi2015}. Furthermore, the slope of the $J$ band PL relation in Reticulum is $2\sigma$ shallower than the slope of the calibrator PLZ relation. Therefore, we adopt the average distance modulus based on theoretical and empirical calibrations in the $HK_s$-bands as the final distance to Reticulum, $\mu = 18.472\pm0.035$~mag, corresponding to a distance of $49.48\pm0.80$~kpc. The error budget includes statistical errors and systematic uncertainties computed by combining in quadrature the zero-point uncertainties of the calibrating PLZ relations ($\sim0.02$ mag) together with the effect of a potential $0.1$ dex uncertainty in the adopted mean metallicity or metallicity scale ($\sim0.02$ mag). The contribution from reddening uncertainties is negligible ($<0.01$ mag) due to  low line-of-sight extinction and less sensitivity of NIR magnitudes on reddening.

\section{Summary}
\label{sec:discuss}

In this first paper in a series on old globular clusters in the LMC, we presented homogeneous time-series observations of Reticulum for the first time in three NIR filters simultaneously. We focussed on 32 RRL variables in Reticulum for which multi-epoch observations obtained with the Flamingos-2 instrument on the Gemini South Telescope, provided well-sampled $JHK_s$ light curves.
NIR light curve template fitting allowed us to determine  accurate intensity-averaged magnitudes and pulsation properties for 22 RRab, 4 RRc, and 6 RRd variables. The RRL stars in Reticulum exhibit tight NIR PL relations with low intrinsic scatter that is comparable to Galactic clusters and nearby dwarf spheroidal galaxies. However, the slopes of PL relations in particular in the $J$-band are systematically shallower than those in Galactic clusters, possibly due to contributions from low-statistics and a narrow range of RRL period distributions. 

Using the calibrated PLZ relations of \citet{bhardwaj2023} in NIR bands, together with the mean-metallicity of the cluster, we determine a true distance modulus of $\mu_0 = 18.472 \pm 0.035$ mag to Reticulum. Given that the calibrator relations utilize metallicities of globular clusters in the \citet{carretta2009} scale, we rely on converting literature metallicities in the same scale using using empirical relations. Nevertheless, the resulting distance is among the most precise independent measurements for Reticulum and is in excellent agreement with the geometric distance to the LMC based on eclipsing binaries \citep{piet2019}. LMC, the host galaxy of Reticulum, is one of the key anchors for the cosmic distance scale. However, there are complications due to its depth, range of metallicities and reddening spread of field stars. With precise and accurate distance determination for Reticulum along with its low-reddening and sparse population, we are aiming towards establishing Reticulum and LMC GCs as potential alternate anchors for the population II distance ladder.

%% Please use the acknowledgment and contribution environments. This will 
%% be anonomyized when the "anonymous" style option is used. 
\begin{acknowledgments}
We thank the anonymous referee for useful comments that helped improve the manuscript. AB thanks the funding from the Anusandhan National Research Foundation (ANRF) under the Prime Minister Early Career Research Grant scheme (ANRF/ECRG/2024/000675/PMS). This research was supported by the International Space Science Institute (ISSI) in Bern/Beijing through ISSI/ISSI-BJ International Team project ID $\#$24-603 – ``EXPANDING Universe'' (EXploiting Precision AstroNomical Distance INdicators in the Gaia Universe). G.D.S. acknowledges financial support from Gaia DPAC funds through INAF and the Agenzia Spaziale Italiana (ASI; contract 2025-10-HH.0, PI: M.G. Lattanzi), the PRIN MUR 2022 project "Early Formation and Evolution of Bulge and Halo" (EFEBHO; grant 2022ARWP9C, PI: M. Marconi), and the Istituto Nazionale di Fisica Nucleare (INFN), Naples Section, through the specific initiatives QGSKY and Moonlight2. This research was supported by the Munich Institute for Astro-, Particle and BioPhysics (MIAPbP) which is funded by the Deutsche Forschungsgemeinschaft (DFG, German Research Foundation) under Germany´s Excellence Strategy – EXC-2094 – 390783311.
\end{acknowledgments}

\begin{contribution}

AB developed this research project and was responsible for the data analysis and writing the manuscript. SD and MRE contributed to the formal analysis and validation, and all authors edited the final version of the manuscript. 

\end{contribution}

\facilities{NOIRLab (Gemini South Telescope -- Flamingos2 imager)}

\software{\texttt{The IDL Astronomy User's Library} \citep{landsman1993}, \texttt{Astropy} \citep{astropy2013, astropy2018, astropy2022}}

%\appendix

%\section{Light curves of Long-Period Variables in globular clusters}
%\label{sec:supp_fig}

\bibliographystyle{aasjournalv7}
\bibliography{mybib_final.bib}



\end{document}